\documentclass[runningheads]{llncs}
\usepackage{graphicx}
\usepackage{subcaption}
\begin{document}
\title{It's more than just money: The real-world harms from ransomware attacks}
\titlerunning{Harms from ransomware attacks}

 \author{Nandita~Pattnaik\inst{1} \and
 Jason R.C. Nurse\inst{1,2}* \and
 Sarah Turner\inst{1} \and 
 Gareth Mott \inst{1}  \and
 \\Jamie MacColl \inst{2}  \and 
 Pia Huesch \inst{2} \and
James Sullivan \inst{2}}

\authorrunning{Pattnaik et al.}

 \institute{Institute of Cyber Security for Society (iCSS) \& School of Computing, \\University of Kent, UK  \and
 Royal United Services Institute (RUSI), UK\\
  \email{J.R.C.Nurse@kent.ac.uk}}

\maketitle              % typeset the header of the contribution
\begin{abstract}
As cyber-attacks continue to increase in frequency and sophistication, organisations must be better prepared to face the reality of an incident. Any organisational plan that intends to be successful at managing security risks must clearly understand the harm (i.e., negative impact) and the various parties affected in the aftermath of an attack. To this end, this article conducts a novel exploration into the multitude of real-world harms that can arise from cyber-attacks, with a particular focus on ransomware incidents given their current prominence. This exploration also leads to the proposal of a new, robust methodology for modelling harms from such incidents. We draw on publicly-available case data on high-profile ransomware incidents to examine the types of harm that emerge at various stages after a ransomware attack and how harms (e.g., an offline enterprise server) may trigger other negative, potentially more substantial impacts for stakeholders (e.g., the inability for a customer to access their social welfare benefits or bank account). Prominent findings from our analysis include the identification of a notable set of social/human harms beyond the business itself (and beyond the financial payment of a ransom) and a complex web of harms that emerge after attacks regardless of the industry sector. We also observed that deciphering the full extent and sequence of harms can be a challenging undertaking because of the lack of complete data available. This paper consequently argues for more transparency on ransomware harms, as it would lead to a better understanding of the realities of these incidents to the benefit of organisations and society more generally.

\end{abstract}

\section{Introduction and Background}

The volume of ransomware attacks --- i.e., malware-based cyber-attacks characterised by blocking access to a device or/and encrypting valuable data~\cite{Yamany2022RansomwareDefinition} --- is constantly increasing, with some reports finding that infections in businesses worldwide are as high as 71\%~\cite{Statista2023RansomwareVictim}. The UK's National Cyber Security Centre highlight this significance by defining ransomware as the most acute threat faced by organisations today~\cite{NCSC2022AnnualReview}. While there have been several articles and reports reflecting on ransomware, its nature, attack patterns, and mitigation strategies~\cite{Oz2022RansomwareSurvey}, there is much less research on the actual negative impacts that can result from these incidents. We characterise such negative impacts using the term \textit{harms}; this is similar to approaches taken by existing research~\cite{agrafiotis2018taxonomy}. Understanding harms from cyber-attacks is vital for a plethora of reasons, especially given their relevance in preparing for the consequences of attacks in the future. As argued by current literature, irrespective of an organisation's threat-driven or impact-driven risk assessment, the limitation of an incomplete understanding of the potential harms and the relationship between those harms can lead to the selection and deployment of inappropriate risk treatments and controls~\cite{agrafiotis2018taxonomy}.

This paper contributes to the field by critically examining the multitude of harms that can arise from cyber-attacks, with a focus upon the present threat of ransomware. We also propose a new methodology by which such incidents and their harms can be comprehensively modelled. Our research makes the point that researchers, businesses and policymakers must go beyond the current focus on financial harms (e.g., payment of ransoms, cost of recovery or cyber insurance claim amounts) to examine all types of real-world harm that can result (e.g., human, physical, social) and how these harms may influence or trigger each other. Ransomware poses a unique case study considering its prominence and ability to cripple unprepared organisations (e.g., UK's NHS and WannaCry~\cite{NHS2018WannaCry}). 

While existing research on ransomware harms and impacts is limited, there are some key articles worthy of review.  
By empirically studying a dataset of 453 ransomware data investigation reports, Meurs et al. reported on specific factors contributing to the ransom requested, the likelihood of ransom payment and their influence on the financial losses~\cite{meurs2022RansomwareFinancial}. They conducted a detailed statistical analysis to present several factors (such as the ransom paid, the revenue of the victims and the use of RaaS (Ransomware-as-a-Service) by an attacker) which were seen to be statistically significant determinants of the financial losses reported. 
Wilner et al., on the other hand, commented on the wider international, political, intelligence and diplomatic ramifications of ransomware by analysing several ransomware cases~\cite{Wilner2019RansomwareSocial}. This is a pertinent example of research into the non-financial and international impacts of such attacks. While these studies generally align with our work, Wilner et al. do not discuss the individual harms that might originate from various ransomware attacks and Meurs et al.'s analysis was focused on factors that contribute to financial harm; rather than an a reflection on differing types of harm.

On the broader concept of harms from cyber-attacks (i.e., not only ransomware), Agrafiotis et al. introduced a taxonomy of harm consisting of five major harm types, namely Physical/Digital, Economic, Psychological, Reputational and Social and societal harms~\cite{agrafiotis2018taxonomy}. This taxonomy was created using a mixed approach of deductive and inductive analysis and based on publicly-available organisational harm data, harm-related literature, and public databases. This enumeration and modelling of harm is one of the closest to our work and while it does not focus on ransomware nor a detailed modelling of harms from attack cases, it can inform our study.
Recent related research has also examined the nature of losses from cyber-related events across different risk categories and business sectors~\cite{Shevchenko2023Losses}. They used a comprehensive database of cyber-loss data over 12 years from 2008–2020, affecting 49,496 organisations across 20 business sectors. That work highlighted the heavy-tailed nature of cyber risks by analysing both the frequency and severity of losses from cyber events. This financial emphasis is clearly relevant to the research and business community but, as mentioned in the previous paragraph, again it risks, not capturing the full range of negative impacts or intangible costs from cyber-attacks.

Studies, particularly Axon et al., have sought to complement existing research by using cyber insurance claims data to build harm-propagation trees that can enhance the understanding of the harms and links between harms after cyber-attacks~\cite{Axon2019}. The graph output from their study is a valuable tool for defining the frequency of each harm's occurrence and also the strength of the relations
between harms. Our research is similar though we benefit from a wider pool of data than what is available from insurance claims. Insurance forms also arguably prioritise harms with a financial component and therefore we expect our study to be more comprehensive in its definition and modelling of harms.  

To address the gap in existing literature related to the definition and understanding of harms from ransomware attacks, we conducted a data-driven, sociotechnical research study. Specifically, we used publicly available data to analyse eight different ransomware incidents and enumerated the harms and harm relations (i.e., which harms lead to other harms) that emerged. These incidents were investigated through the construction of a series of ransomware harm models enumerating the relevant data. In addition to providing an improved appreciation of the long tail of harms after a ransomware incident, we posit that the modelling methodology proposed and these models themselves are significant for two reasons. First, they provide businesses with data that is necessary for effectively implementing risk controls within their organisations. That is, they encourage consideration of harms beyond initial server compromise or loss of data to wider harms that negatively affect the business and its stakeholders. Secondly, the methodology and resulting models explicitly highlight the wide nature of harms to researchers studying cyber-attacks and policymakers responsible for protecting an increasingly digital society.

\section{Methodology}
\label{Methodology}

\subsection{Definition and scope}
 
The first step in our research process was to define its parameters and scope. Harm, as described earlier, is any negative impact that can occur from a cyber incident; a description adapted from existing work~\cite{agrafiotis2018taxonomy}. By their nature, harms can be vast and can transpire immediately (e.g., a compromised and inaccessible server) or in the longer term (e.g., a regulatory fine years after suffering a data breach). To facilitate a structured extraction and analysis of harms emerging from ransomware attacks, we decided to adopt an existing harm taxonomy~\cite{agrafiotis2018taxonomy}. This taxonomy provided an initial list of validated types of harm that could act as a foundation for our work. In terms of the cases scoped for data gathering, our choice of scenarios was informed by two factors: well-publicised or high-profile ransomware cases that took place at least three months prior (i.e., before December 2022), and sectors regularly impacted by ransomware attacks. The first factor was important because such cases would have more extensive reports and media coverage for us to draw on, and we would also be able to track harms over a longer period of time (i.e., not only immediately after the incident). The second factor was necessary to understand the extent and type of harms initiated and propagated by the frequent attacks in certain specific sectors. 

In total, we selected and assessed eight incidents: (1) NHS, UK -- WannaCry, 2017, (2) Health Service Executive (HSE), Ireland -- Conti, 2021, (3) Hackney Council, UK -- Pysa, 2020, (4) Atlanta City government, US -- SamSam, 2018, (5) Colonial Pipeline, US -- Darkside, 2021, (6) Travelex, UK -- REvil, 2020, (7) UK Schools -- Vice Society, 2022, and (8) Los Angeles Unified School District (LAUSD), US -- Vice Society, 2022. These incidents represent significant ransomware attacks across highly impacted sectors such as healthcare, energy, government and finance. Due to space constraints, we will report on only two cases, the HSE Conti attack and the Hackney Council Pysa attack. These cases were specially chosen because they present relevant exemplars of the multitude of harms that can emerge from ransomware incidents and aptly capture some of the central themes arising from the other cases.

\subsection{Collection and analysis of cases}
For each case, a web search using the name of the organisation and the ransomware attack/group was used to source a diverse range of relevant articles, including audit investigation reports, where available, alongside newspaper articles, media reports and academic literature. Once collected, we extracted insights pertinent to the harms that occurred and the relations between harms (i.e., how a harm may lead to, or trigger, another harm). The range of sources amassed for each incident was crucial in creating a more complete picture of each attack. To guide our harm annotation and extraction process, four rules were followed. 

\begin{itemize}

\item \textbf{Rule 1 (R1)}: \textit{In cases where the harms---as guided by the harm taxonomy~\cite{agrafiotis2018taxonomy}---from the attack were directly stated in the article's text, these should be recorded and extracted as harms emerging within the case being studied.}

\item \textbf{Rule 2 (R2)}: \textit{When an article's text does not precisely use a harm in its writing, but its connotation indicated a strong affinity towards a particular harm, annotate the paragraph/sentence as close as possible to the above-indicated meaning.} 
To demonstrate this point we use the following example excerpt taken from one of the case reports of Hackney Council~\cite{WSJ2021Hackney}. For the excerpt, ``some residents in the borough are still waiting for payments for various benefits'' the harm in this situation was annotated as ``Financial loss to residents'' (as this pertains to residents not being able to receive payments that would allow them to access the welfare benefits they are entitled to).

\item \textbf{Rule 3 (R3)}: \textit{To more clearly articulate and present the harms discovered, similar harms should be placed into representative groups}. 
The following is an example taken from the HSE  case~\cite{HSE2021Review}: 
``In the community, primary care staff were unable to access patient appointment lists or contact details, patient history, treatment plans, x-ray facilities or monitoring of instrument sterilisation tracking.'' This paragraph was recorded and noted as a single harm, namely ``Loss/unavailable clinical data'', so as to capture the clinical nature of the data impacted, without creating new harms for each of the individual data types (e.g., patient history, treatment plans). 

\item \textbf{Rule 4 (R4)}: \textit{Once a harm was identified, we also reviewed the article's text for any other harms that occurred as a result of (i.e., were triggered by) that initial harm, and these were recorded and extracted as harm relations.} For example, consider the following text, ``Inability to use HSE email accounts led to a delay in the General Register Office process leading to delays in child benefit payments for new births'' that was extracted from the independent post-incident review of the Conti ransomware attack on HSE~\cite{HSE2021Review}. This led to the definition of the harm relations ``Unavailable non-clinical system $\rightarrow$ Disrupted/delayed non-clinical services'' and ``Disrupted/delayed non-clinical services $\rightarrow$  Delayed child benefit payment''.  

\end{itemize}
 
Each article, in turn, was then analysed and coded by two researchers separately according to these rules. Both harm and harm relations were stored in Mendeley\footnote{https://www.mendeley.com/} with all relevant document texts annotated as denoted above. Rules were essential for reducing subjectivity in the harm recording and extraction process, and to further validate the harms extracted, annotated texts from the articles were discussed across the author team. A representative sample of harms and relations from texts were also validated by a group of four researchers to settle any differences and produce an agreed set of harm and harm relations.

\subsection{Harm model design}

The primary aim of this research is to examine, and provide a methodology for highlighting, the multitude of harms that can arise from ransomware attacks, thereby providing an evidence base for an increased acknowledgement and understanding of these harms. To support this aim and to portray harms and their relationships visually, we constructed a series of harm models (one for each case) using the harm list and harm relations developed earlier (and based on the rules above). Models are a well-known technique to characterise complex real-world phenomena and have also been applied to explain harms in prior literature~\cite{agrafiotis2018taxonomy,Axon2019}. 

We depict the harm model as a non-weighted directed graph $G = (V, E)$, where each node $u \in V$ represents an observed harm from the ransomware incident and each directed edge $(p, q) \in E$ indicates that a relationship exists between the two harm nodes $p$ and $q$ (i.e., that harm $p$ has been observed as causing or otherwise leading to harm $q$). Modelling harms as a directed graph also has other advantages --- e.g., in detecting and preventing the possible propagation of harms, a key task for risk managers --- as will be discussed later in this paper. This design methodology is also central to our contribution.

\section{Results}

We structure this section by first presenting two of the cases that were modelled  to provide further insights into the cases studied, the harms and how they emerged, and the models designed (using the process above) to demonstrate harms and their relations. The section then progresses to report on key observations and findings from the complete set of eight case studies. These observations are central to our research contribution as they present unique findings related to the wider understanding of harm from ransomware attacks. 

\subsection{Hackney Council, UK, 2020 attack by Pysa Ransomware}

\begin{figure}[!h]
    \centering
    \includegraphics[width=\linewidth]{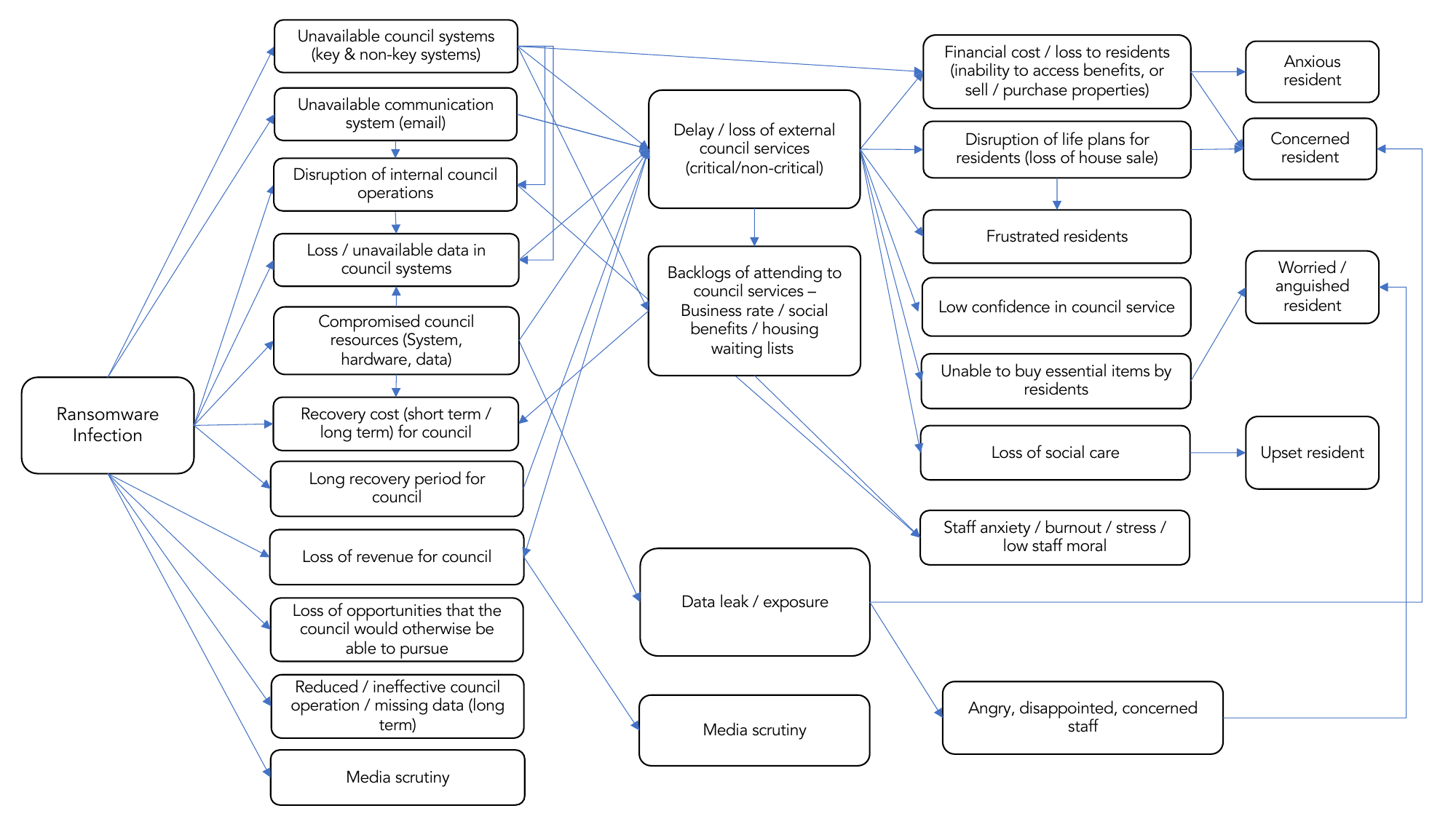}
    \caption{Harm model for Hackney Council, UK -- Pysa, 2020}
    \label{fig:ModelHackney}
\end{figure}

The first harm model to be presented covers the attack by Pysa Ransomware on Hackney Council. In October 2020, Hackney Council, a local authority within Greater London in the UK, came under attack by the Pysa ransomware group. The attack compromised essential council resources making them inaccessible. It consequently brought most of the council's operations to a standstill~\cite{Wired2023Hackney}. The various harms discovered in this case can be seen in Fig.~\ref{fig:ModelHackney}. As depicted in the figure, examples of harms that emerged in the aftermath of the ransomware attack included \textit{Compromised council resources}, \textit{Unavailable council systems}, \textit{Loss/unavailable data in council system} and \textit{Disruption of internal council operations}. This led to the shutting down of several of the council's key external-facing services, such as the social benefits system and social care services. 

The inability to serve residents living in the local authority area was exacerbated with the passing of time (as long as two years)~\cite{WSJ2021Hackney}, resulting in an expensive clean-up and recovery cost of nearly £12 million, a huge backlog of work and subsequent leak of data publicly~\cite{BBCNews2021Hackney2}. 
We can visualise some of these downstream harms in the figure as \textit{Delay/loss of external council services}, \textit{Recovery costs}, \textit{Backlog of attending to council services} and \textit{Data leak/exposure} in the Hackney Council harm model. To explain this in the context of our harm model design notation, when \textit{Data leak or exposure} ($p$) is connected to \textit{Concerned residents} ($q$), it demonstrates the fact that that there is a relationship between these two nodes and it has been observed in the data that the harm node \textit{Data leak or exposure} might lead to \textit{Concerned residents}.

The non-functioning of crucial services such as social care services and benefit payments affected thousands of local residents whose daily lives were dependent on them. Harms affecting both staff and individual residents within the local authority area can be seen towards the right of the model, portraying both the loss of facilities i.e., \textit{Unable to buy essential items}, \textit{Disruption of life plans} and various psychological harms, e.g., \textit{Concerned resident} and \textit{Worried resident}.

\subsection{HSE, Ireland, 2021 attack by Conti Ransomware}

\begin{figure}[!htb]
    \centering
    \includegraphics[width=\linewidth]{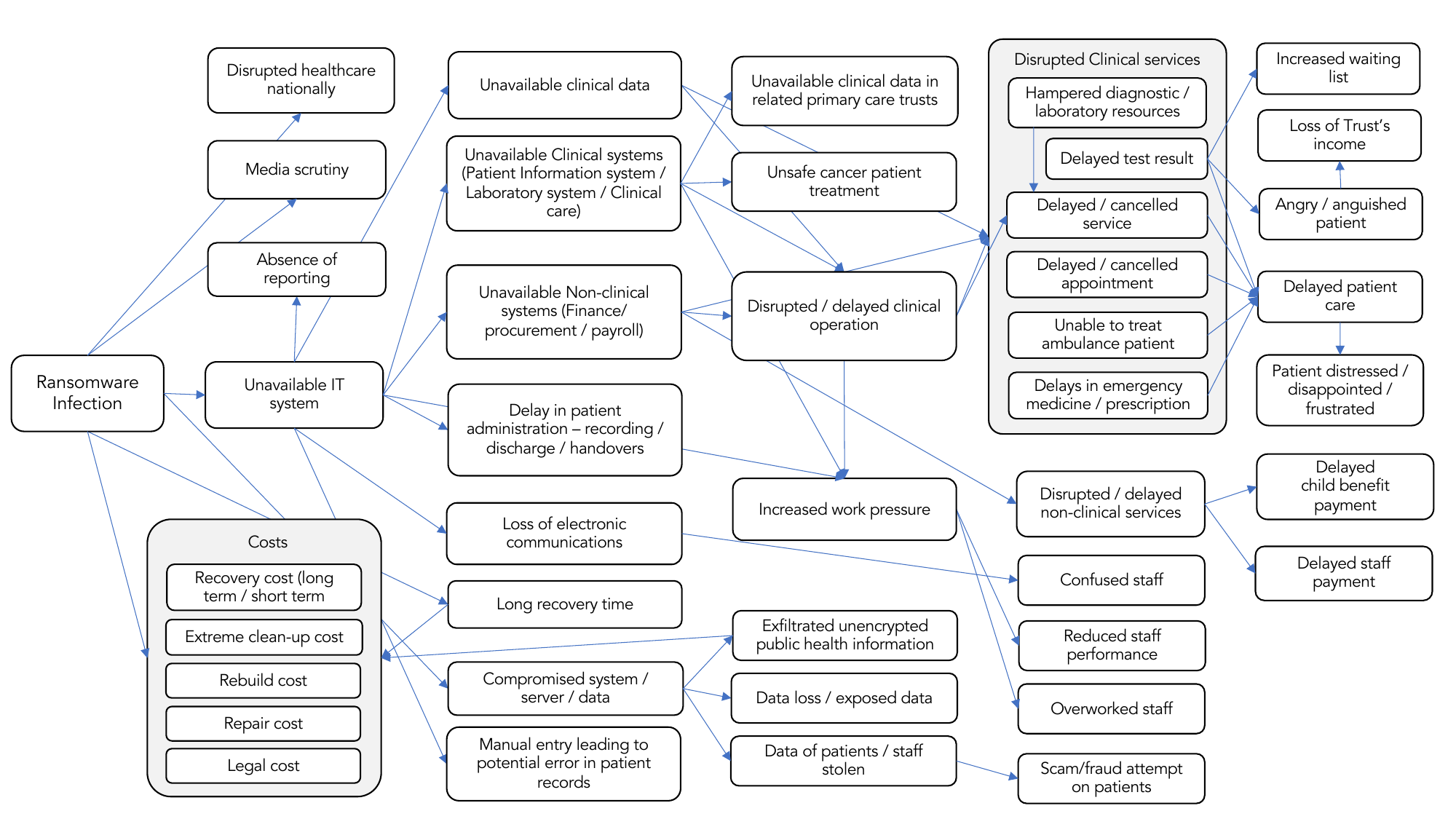}
    \caption{Harm model for HSE, Ireland -- Conti, 2021}
    \label{fig:ModelHSE}
\end{figure}

The HSE, Ireland's biggest public sector employer, was hit by a ransomware attack in 2021 leading to the closure of HSE's 4,000 locations, supporting 54 acute trusts and 70,000 devices~\cite{HSE2021Review}. The harms resulting from the attack to the hospitals, patients, clinical and non-clinical staff, and all third-party users of the hospital system were long-lasting, widespread, and devastating. Specifically, the ransomware infection led to an immediate shut down of all hospital IT-driven clinical facilities resulting in harms including~\textit{Unavailable clinical system (Patient information system/Laboratory system/Clinical care)}, \textit{Unavailable non-clinical systems}, \textit{Unavailable clinical data} to mention a few~\cite{HSE2021Review}. The respective harm model is shown in Fig.~\ref{fig:ModelHSE}. The aforementioned harms triggered a host of subsequent harms for patients and staff such as \textit{Patient} being \textit{distress/disappointed/Frustrated} or \textit{Confused staff} and \textit{Reduced staff performance}. The compromised system also led to the leak of sensitive patient data (i.e., \textit{Data loss/exposed})~\cite{BankInfoSecurity2023HSE}. In Fig.~\ref{fig:ModelHSE} we have also grouped two sets of related harms, i.e., Costs and Disrupted clinical services, primarily for ease of visualisation. This does, however, also have the benefit of showing how a single harm can led to various others; for instance, \textit{Disrupted/delayed clinical operation} causing a host of \textit{Disrupted clinical services}. 

\subsection{Observations from case analysis and modelling}

The process of identifying and modelling harms and their relationships by drawing on publicly-available data provided us with substantial insight into the real-world consequences of ransomware attacks. There are several salient observations that can be made from this research.

Ransomware attacks can result in a significant and diverse set of harms substantially beyond financial impacts. This point emerged clearly from our case studies. Physical/digital harm was one of the most common harm types and presented in every case we analysed; this was undoubtedly because ransomware attacks primarily aim to encrypt/block digital resources as a prerequisite to demanding a ransom. More specifically, the assessed ransomware cases depicted the physical/digital harms of \textit{Unavailability of resources}, which subsequently can led to \textit{Disruptions of internal operations} and likely then to \textit{Disruption of (external) services}. Another example of a common digital harm was the \textit{Stolen/exposed data} as seen in the ViceSociety attack on Los Angeles Unified School Districts (LAUSD)~\cite{WIRED2022RansomViceSociety} where the hackers allegedly stole 500GB of data from LAUSD. 

Economic harms are the other set of common harms that result from a ransomware attack. These manifested in many different forms in the cases observed. For instance, we noted \textit{Ransom costs} in the SamSam attack on the Atlanta government~\cite{WIRED2023RansomAtlanta}, \textit{Recovery costs} in the REvil attack on Travelex~\cite{WSJ2020TravelexRansomPaid}, and \textit{Clean-up costs} in the Conti attack on HSE. Apart from the aforementioned harms, which almost always receive more attention in public, there are, of course, other sets of harms, i.e. psychological and societal harms, which are equally important and, more often than not, materialised as a consequence of the harms above. This reiterates the fact that ransomware harms are more than just the financial and monetary impact. We could see various examples of such harm presented in our data. For example, there were psychological harms in \textit{Frustrated/upset employees} in the REvil attack on Travelex~\cite{BBC2020Travelx2} and \textit{Stressed staff} in the WannaCry attack on the NHS~\cite{Wheeler2022NHS}. At a societal level, \textit{Increased fuel prices nationally} due to the Darkside attack on Colonial Pipeline~\cite{CNN2021Colonial} and \textit{Disrupted healthcare nationally} after the Conti attack on HSE, were also apparent. 

One difference to prior research on wider cyber-attacks that was identified from our analysis of cases was less coverage of reputational harms (i.e., negative external impressions on the impacted organisation) in some ransomware attacks. This is surprising given that cyber-attacks usually result in comprehensive negative impacts on the reputation of the breached organisation~\cite{nurse2020framework}. Our research did find that there was significant media attention and scrutiny placed on the compromised organisations as a result of the attack (undoubtedly due to their public-facing nature, the large-scale impact, and the money spent on the response). However, there was never a clear link to wholesale damage to the organisation's image. While it is out of the scope of this research to determine the reasons for this, some possibilities include sympathy for the victim organisation given, for instance, the well-resourced nature of these threat actors~\cite{govuk2023ransom}, or a feature of specific sectors (e.g., the lack of coverage was salient with government entities in particular). Alternatively, this may also represent a limitation in the publicly accessible data comprising our dataset. 

In assessing the range of harms, another salient observation was the absence of appropriate methods to formally capture and record the full set of 
harms that may transpire. For instance, in literature~\cite{Wheeler2022NHS} covering the NHS WannaCry attack, it is explained that a ransomware-related death would currently be impossible to formally report --- and thus officially recognise a harm --- as there is no code to input into a hospital form for that particular incidence, i.e. death due to a cyber/ransomware incident. This is one example, but we observed similar incidents where there were no protocols to properly capture/record, and therefore acknowledge or report on, the extent of harms from a ransomware attack. 

Grounding our analysis and modelling in prior work (e.g., the harm taxonomy~\cite{agrafiotis2018taxonomy}) proved particularly useful as it enabled us to define a structured set of harms that was also closely linked to the actual data. For example, the generic physical harm of ``Damaged or Unavailable'' was adapted to ``Unavailable council system (Key \& non-key system)'', ``Loss/unavailable data in council systems", and ``Unavailable communication systems'' in the Hackney case study. Similarly, we were able to present a level of granularity in our models that expanded beyond general recovery costs to different types of specific costs (recovery, rebuild, clean up, opportunity, etc.) that featured in mitigation, recovery and rebuilding in the aftermath of the incident. These further signified the more complicated and detailed nature of the cost involved. In general therefore, our work acts to further validate that taxonomy and exemplify how it can be applied and extended.

The analysis conducted also discovered a complex web of interconnected harms caused by ransomware attacks; this is aptly depicted in the harm models. The rules were especially useful here as they allowed us to assess the cases in depth and consider the various chains of events that arose which then depicted different harms and sequences of harms. The Colonial Pipeline ransomware attack by the Darkside group provided one such example. For instance, the shutdown of major gasoline pipelines (i.e., \textit{Disruption to gasoline supply}) led to a reduction in gasoline availability (i.e., \textit{Unavailable gasoline resources}), which caused anxiety and panic buying by consumers (i.e., \textit{Anxious and panicked consumers}) and also resulted in a spike in gas prices. The situation became life-threatening when a car carrying four cans of gasoline burst into flames; although no-one was killed, this also  resulted in more physical damage (i.e., \textit{Destroyed property} harm)~\cite{WashingtonPost2021Ransom,TehTarget2022Colonial}.    

One point to further explore from the modelling process was that ransomware attacks initially impact the technology and systems, but there is often subsequent harm affecting individuals. This builds on our earlier observation on psychological harms and their common occurrence across attacks. If we reflect on the Hackney Council case, for example, six out of the rightmost eight leaf harm nodes represent either harm to the residents of the council or harm to the staff working there. Reviewing the models broadly, harm relations often start with some digital/physical or economic harm and ultimately lead to harm to individuals, as depicted in the HSE model \textit{Unavailable clinical system} leads to \textit{Disrupted clinical services} which in turn leads to \textit{Angry/anguished patient}, i.e., harm to the individual. This demonstrates the long tail of harms and highlights the social/human harms organisations might overlook as they are more difficult to assess, measure and accommodate.

\begin{figure}[!htb]
    \centering
    \includegraphics[width=.9\linewidth]{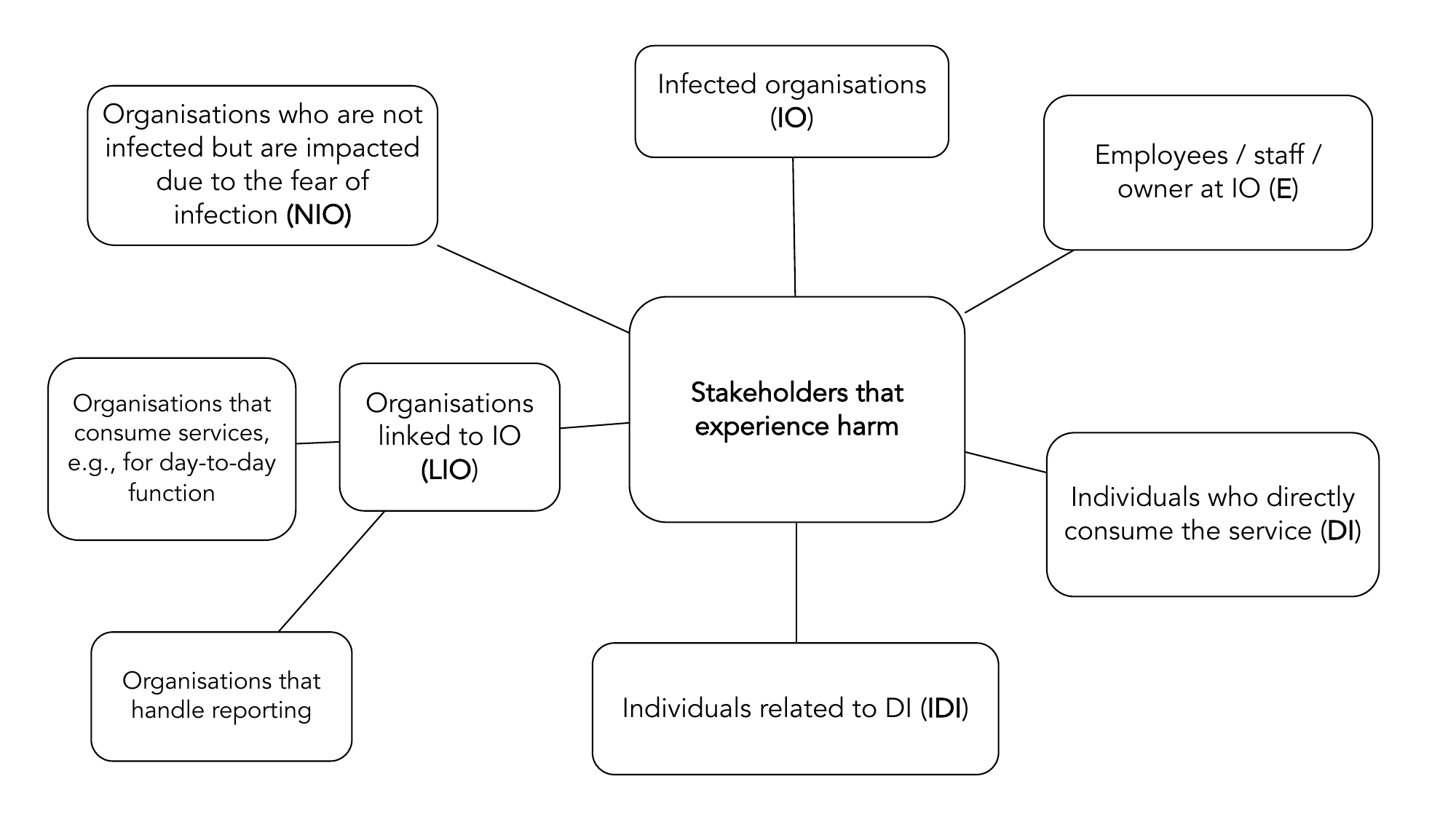}
    \caption{Types of stakeholders that can experience harm}
    \label{fig:stakeholders}
\end{figure}

\begin{figure}[!htb]
    \centering
    \includegraphics[width=1\linewidth]{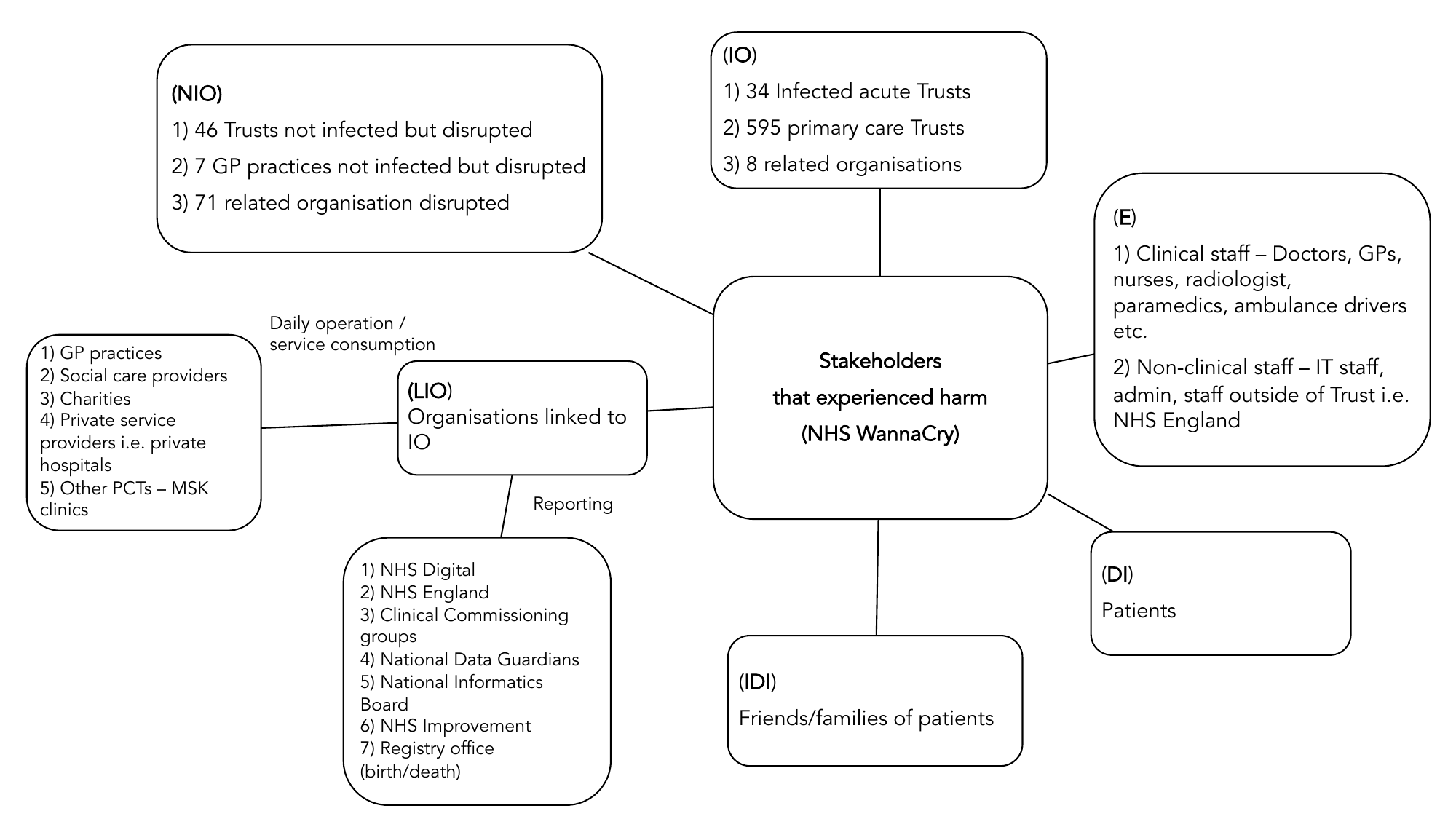}
    \caption{Stakeholders that experienced harm during the NHS WannaCry ransomware attack}
    \label{fig:stakeholders-ex}
\end{figure}

To complement the observations above and summarise the various stakeholders that can be harmed by ransomware attacks as identified from our analysis, we present Fig.~\ref{fig:stakeholders}. This spotlights the infected organisation but also the several other entities and individuals likely to experience some form of harm. We take this opportunity to provide some insight into another case, namely the NHS WannaCry attack, and also present the stakeholders that experienced harm in Fig.~\ref{fig:stakeholders-ex}. Having a clear idea of who might be the affected parties, what harms affect them, and what triggers those harms could put the businesses and policymakers in a better position to respond to attacks and draft appropriate policies. 


A final notable observation was that the \textit{fear} of ransomware attacks could also prompt the same types of harm in uninfected organisations with a link to (e.g., in the same supply chain or enterprise context as) the infected one, as in infected or compromised ones. This was witnessed in the NHS case where 46 non-infected Trusts had to shut down their operations in fear of infection and therefore experienced the same harms as the infected Trusts. The situation was even worse for some of these groups because due to the attack they were unable to get online to then execute the kill switch needed to stop the attack~\cite{NHS2018WannaCry}.

\section{Discussion and conclusion}

\subsection{Discussions}
This paper contributed to existing work by investigating, and providing a methodology to explore, the plethora of real-world harms emerging from ransomware attacks, thereby directly informing the sociotechnical evidence base for researchers, businesses and policymakers. We engaged in a study of these harms by reviewing well-documented cases of ransomware attacks, creating models to understand the presence and relations between harms, and critically reflecting on these to extract a set of pertinent observations of importance to the wider community. 

To comment more broadly on our findings, the nature and extent of harms was extensive. We were able to identify harms that are rarely acknowledged in research or industry and link these directly to the threat of ransomware attacks. We also noticed each set of harm triggered a chain of harms i.e., infections lead to unavailable data which leads to disruption of services, which can result in direct harm to employees or other individuals. This detailed identification of harms and also the harm relations provide valuable new information for organisations and policymakers seeking to implement measures that limit the harm caused by ransomware attacks. 

Organisations can draw on our work's insights to: (a) improve the accuracy of their risk assessments and subsequent risk treatments because they would be able to incorporate a more complete set of harms that can emerge from ransomware-related risks; and (b) set up appropriate business continuity processes and incident response plans in preparation for a ransomware attack. Moreover, models such as these might serve as a sector-specific blueprint of harm propagation in case of ransomware attacks on certain sectors (e.g., healthcare or education) and help affected parties, including governments (who need to understand the harms of ransomware on critical infrastructure, healthcare, etc.), to plan preemptively. Generally, this modelling process also provides methodological guidance for policymakers in identifying the type and trajectory of ransomware harms which can then be used to develop more formalised cyber harm models.

Although temporal factors are not represented in these models, our analysis indicated that digital/physical and some economic harms are often experienced in the short-term period after/during an attack. Psychological harms on the other hand might be immediate or delayed depending on the nature of the service affected. This discussion around harms and their sequence is an interesting one as a better understanding of sequence provides an opportunity for remediation and preventing further harms. One challenge for organisations will undoubtedly be how far downstream in a harm model to consider and what is appropriate to include when assessing a cyber risk (inclusive of its impact).

\subsection{Limitations and future work}
It is challenging to understand the full extent of harm resulting from ransomware attacks. We used published reports, articles and literature as the basis for our study given the richness of information it presented. This information, however, is likely to be incomplete as there is almost certainly information that was withheld by the organisation or was not covered in the publicly-accessible reports that we were able to source. A relevant example is the HSE case and its harm model. This model was one of the richest in our study but this was undoubtedly influenced by the fact that there was an official audit report that was publicly released; only the NHS attack also had a similar public report. This further highlights some of the issues of exploring the harms of ransomware, or any cyber-attack; that is, a complete understanding may not be feasible even years after an attack. 

Another related factor is the sequence of harms and harm relations identified. Our work aimed to represent the relationships present in the data and did not prejudge or rearrange stated sequences. As such, if it were to transpire that a relation was not accurate or overly simplistic, this would impact our work. We did attempt to address this issue via triangulating harms and relations across sources, however this is still a potential weakness. In spite of these potential issues, this research presents one of the few contributions to better understand harms of ransomware attacks, and thereby provide an evidence base beyond financial consequences. 

There are two primary avenues for future research. The first involves expanding the set of cases studied to explore a few sectors in more depth in order to determine whether there are any patterns of harm (especially sequences and harm relations). The health sector is of most interest given the constant stream of attacks, the extensive coverage it tends to attract, and that it may release audit reports into the attack (as seen with HSE and NHS). Understanding patterns of harm is useful as it provides an opportunity to break the spread of harm and thereby limit the stakeholders impacted. The second avenue builds on the first and would seek to encode harm models such that automated analysis of harms across cases --- ours or any others provided by the community --- could be achieved. This would allow a quicker identification of patterns and would also ease uptake for organisations considering integrating our work into their risk analysis methods or policymakers reflecting on sector-wide harms. 

\section*{Acknowledgements}
This research was funded by The Research Institute for Sociotechnical Cyber Security, a collaboration of the UK's Engineering and Physical Sciences Research Council (EPSRC) and the National Cyber Security Centre (NCSC). We also thank Keenan Jones for contributions to the earlier parts of this research.
\bibliographystyle{splncs04}
\bibliography{paper}
\end{document}